\documentclass[pdflatex,sn-apa]{sn-jnl}

\usepackage{graphicx}%
\usepackage{multirow}%
\usepackage{amsmath,amssymb,amsfonts}%
\usepackage{amsthm}%
\usepackage{mathrsfs}%
\usepackage[title]{appendix}%
\usepackage{xcolor}%
\usepackage{textcomp}%
\usepackage{manyfoot}%
\usepackage{booktabs}%
\usepackage{algorithm}%
\usepackage{algorithmicx}%
\usepackage{algpseudocode}%
\usepackage{listings}%

\usepackage{times}
\usepackage{latexsym}
\usepackage[T1]{fontenc}
\usepackage[utf8]{inputenc}
\usepackage{multicol,multirow,makecell}
\usepackage{microtype}
\usepackage{inconsolata}
\usepackage{graphicx}
\usepackage{booktabs}
\usepackage{caption,diagbox}
\usepackage{float}
\usepackage{url}
\usepackage{multirow}

\unnumbered

\begin{document}

\title{The Lovelace Test of Intelligence: Can Humans Recognise and Esteem AI-Generated Art?}

\author[1]{\fnm{Ewelina} \sur{Gajewska}}\email{ewelina.gajewska.dokt@pw.edu.pl}
\affil*[1]{\orgname{Warsaw University of Technology}, \orgaddress{\city{Warsaw}, \country{Poland}}}

\abstract{This study aims to evaluate machine intelligence through artistic creativity by employing a modified version of the Turing Test inspired by Lady Lovelace. It investigates two hypotheses: whether human judges can reliably distinguish AI-generated artworks from human-created ones and whether AI-generated art achieves comparable aesthetic value to human-crafted works. The research contributes to understanding machine creativity and its implications for cognitive science and AI technology. 
Participants with educational backgrounds in cognitive and computer science play the role of interrogators and evaluated whether a set of paintings was AI-generated or human-created. Here, we utilise parallel-paired and viva voce versions of the Turing Test. Additionally, aesthetic evaluations are collected to compare the perceived quality of AI-generated images against human-created art. This dual-method approach allows us to examine human judgment under different testing conditions. 
We find that participants struggle to distinguish between AI-generated and human-created artworks reliably, performing no better than chance under certain conditions. Furthermore, AI-generated art is rated as aesthetically as human-crafted works. Our findings challenge traditional assumptions about human creativity and demonstrate that AI systems can generate outputs that resonate with human sensibilities while meeting the criteria of creative intelligence. 
This study advances the understanding of machine creativity by combining elements of the Turing and Lovelace Tests. Unlike prior studies focused on laypeople or artists, this research examines participants with domain expertise. It also provides a comparative analysis of two distinct testing methodologies—parallel-paired and viva voce—offering new insights into the evaluation of machine intelligence.}


\maketitle

\section{Introduction} 
In 1950, Alan Turing posed the question, ``Can machines think?'' \citep{turing1950computing}, starting a discussion about artificial intelligence, still relevant today. While the question itself was too meaningless to deserve consideration for Turing at that time, the researcher proposed the idea of the imitation game, known later under the name of the Turing Test (TT). 
Turing posited that if a human judge is unable to distinguish between a machine and a human during a conversation, the machine could be considered intelligent. What is more, Turing gave a prediction that by the turn of the century, during a 5-minute human-machine interaction, a lay person would have no more than a 70\% chance of making a human/machine determination \citep{turing1950computing}. 
Failure to achieve the proposed threshold would deem a machine to be intelligent. This percentage reflected Turing's estimation of the difficulty of the test for machines at that time, as he considered AI technology still in its infancy. 
Shortly after Turing's publication, TT started to be viewed as the ultimate test of machine intelligence and is regarded as such nowadays \citep{french2012dusting}.
Later interpretations of TT, however, set a 50\% threshold for the Turing Test, suggesting that equal performance (i.e., a 50/50 distinction) better represents the idea that a machine's ability to imitate human intelligence convincingly should result in equal confusion between human and machine responses \citep{french2000turing,shieber2007turing}.

\begin{figure}[h]
    \centering
    \includegraphics[width=0.65\linewidth]{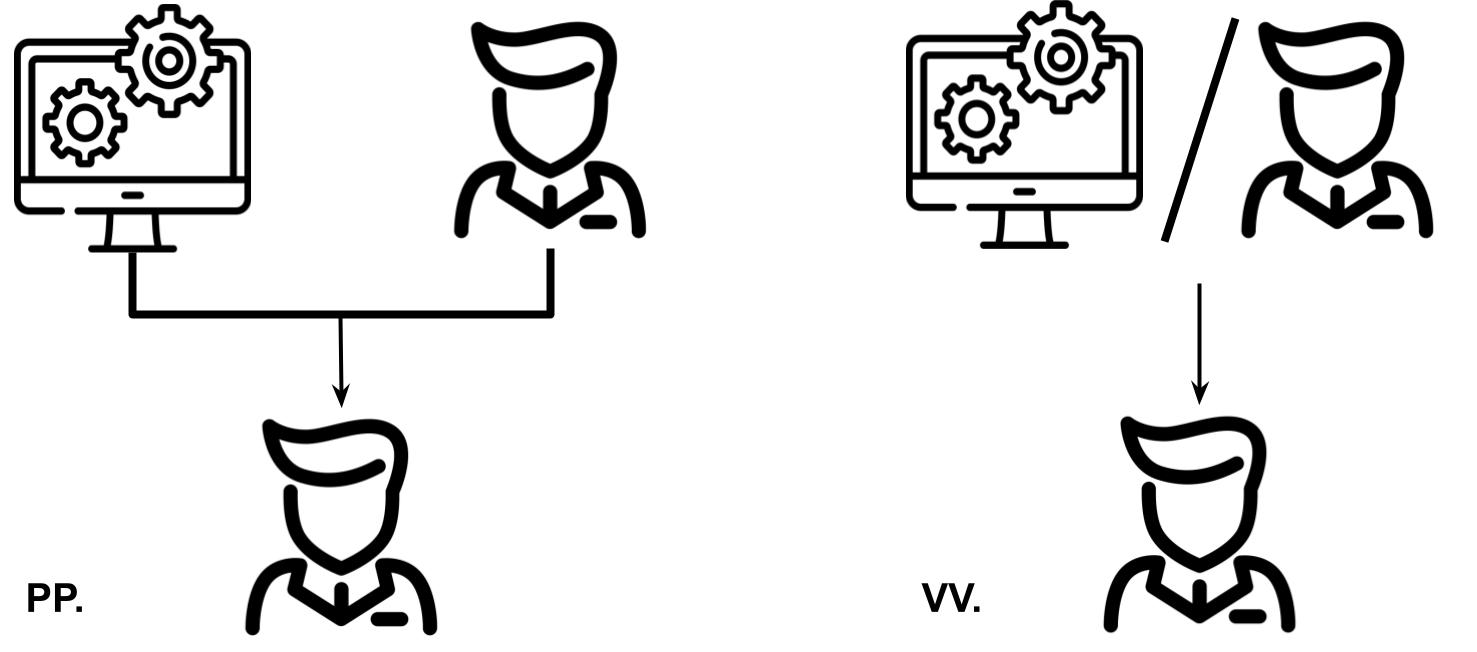}
    \caption{Two versions of the Turing Test: parallel-paired (PP) and viva voce (VV) test.} \label{fig:tt}
\end{figure}

Furthermore, Alan Turing proposed two versions of his imitation game (see Figure \ref{fig:tt}). 
The parallel-paired Turing Test involves three participants: a human judge, a human respondent, and a machine respondent. The judge interacts with both respondents simultaneously and must determine which is human. Studies, such as those conducted during the Loebner Prize competitions, have demonstrated the challenges machines face in mimicking human-like responses convincingly \citep{warwick2016can}. 
The viva voce Turing Test is a simplified version of Turing's original imitation game, involving a one-on-one interaction between a human judge and either a human or a machine. This format allows the judge to engage in a deeper and more focused conversation, probing the entity's reasoning, understanding, and conversational abilities. Unlike the parallel-paired test, which involves simultaneous comparison with a human, the viva voce test isolates the machine's performance, emphasizing its ability to sustain a coherent and meaningful dialogue. 

Research comparing these two formats, such as experiments at Bletchley Park, has highlighted the strengths and weaknesses of each \citep{shah2014fundamental}. The parallel-paired test is often seen as a more rigorous challenge due to the direct comparison with a human, while the viva voce test provides a more focused evaluation of the machine's conversational abilities. Both versions continue to inspire debates about the nature of intelligence and the evolving capabilities of AI systems.

However, as artificial intelligence technologies have advanced, the limitations of the Turing Test have become increasingly apparent. Critics argue that the TT primarily measures a machine's ability to mimic human conversation rather than its capacity for genuine understanding or creativity \citep{browning2024personhood}. This has led to calls for alternative assessments that more accurately reflect the complexities of machine cognition. 
\citet{bringsjord2001creativity} propose one such alternative - the Lovelace Test, named after Ada Lovelace, who famously asserted that machines cannot be deemed intelligent unless they can create something novel and creative. It shifts the focus from conversational mimicry to the ability of artificial agents to produce creative works, thereby providing a more rigorous measure of machine intelligence.

The study contributes to the ongoing discussion on artificial intelligence and creativity by providing empirical evidence regarding the perceptual capabilities of human judges in recognizing AI-generated art. By examining the identification of machine-generated images and their aesthetic evaluations, we aim to deepen our understanding of how it is perceived in relation to human artistry. 
With the combination of the Turing and the Lovelace Test, we investigate two specific hypotheses regarding the topic: first, human judges will perform no better than chance in distinguishing between AI-generated artworks and those created by humans (H1); second, AI-generated images will achieve a level of aesthetic value comparable to that of human works, suggesting that machines can produce art that resonates with human sensibilities (H2). 
To address these hypotheses, we formulate two research questions:
\begin{itemize}
    \item RQ1: Can human judges reliably distinguish between AI-generated and human-created artworks?
    \item RQ2: Is the aesthetic value of AI-generated images comparable to that of human-created images?
\end{itemize}

\section{The Lovelace Test of Intelligence}
\citet{bringsjord2001creativity} argue that original the TT falls short as a test of intelligence because of issues raised by the well-known Chinese Room Argument \citep{searle1980}. 
That is, a machine is designed to mindlessly follow a strict rulebook. 
They argue for a certain epistemic relation between an artificial agent \textit{a}, its output \textit{o}, and the human architect \textit{d}. 
In their proposition of a modified TT (the Lovelace Test) the authors say that artificial agent \textit{a}, designed by \textit{d}, passes the test if: (1) \textit{a} outputs \textit{o}; (2) \textit{o} is the result of processes \textit{a} could repeat; (3) \textit{d} cannot explain how \textit{a} produced \textit{o}.

\citet{riedl2014lovelace} proposes the updated Lovelace Test which is designed to be employed as a test of human-level intelligence for artificial intelligence. 
Although the test seems to be formulated in a way so as to test the creative ability of AI systems, \cite{riedl2014lovelace} argues that this ability requires a wide range of human-level capabilities and, therefore, could be regarded as a test of human-level intelligence for AI systems.
In order to pass this version of the test, four conditions must be fulfilled: (1) artificial agent \textit{a} needs to create an output \textit{o} of type t; (2) output \textit{o} complies with a set of criteria \textit{C} that are expressible in natural language; (3) a human judge \textit{j}, having chosen t and \textit{C}, is satisfied with \textit{o} as a valid instance of t that meets \textit{C}; (4) a human referee \textit{r} determines that the combination of t and \textit{C} as being realistic for an average human.

At the same time, Riedl defines computational creativity as ``the art, science, philosophy, and engineering of computational systems that, by taking on particular responsibilities, exhibit behaviours that unbiased observers would deem to be creative'' \citep[p. 2]{riedl2014lovelace}. The essence of the updated Lovelace Test is the fulfilment of a set of constraints required by a human judge -- if a machine is able to respond to those constraints no worse than an average person, it is compelling evidence of intelligence.

\citet{boden2010turing} proposes two alternative criteria for passing the Turing Test in the domain of machine-generated art.  
``I will take it that for an `artistic' program to pass the TT would be for it to produce artwork which was: (1) indistinguishable from one produced by a human being; and/or (2) was seen as having as much aesthetic value as one produced by a human being'' \citep[p. 409]{boden2010turing}. 
Novel AI systems such as DALL-E 2 developed by OpenAI\footnote{\url{https://openai.com/dall-e-2/}} are able to account for the first objection -- interactivity in the modified TT. DALL-E 2 can create realistic images from a text prompt, i.e. description of a task in natural language. A human judge could, therefore, formulate specific criteria for the image that AI (and a human artist) must produce. Then, if a machine and a human artist are provided with the same instructions (analogous to questions asked by a judge in the original TT) both artworks can be directly compared and evaluated by the judge.

\section{Related Work}
The predominant methodology across studies involves variations of the Visual Turing Test, where participants classify images as AI-generated or human-created. The predominant methodology across studies involves variations of the Visual Turing Test, where participants classify images as AI-generated or human-created. The work of \cite{hong2019artificial} examined the human perception of artwork generated by artificial intelligence (AI). Specifically, the researchers tested how the knowledge about the author's identity (machine vs. human) affects human evaluation of art. Judgements of artistic value were not equivalent between artwork produced by AI and humans -- the latter condition gathered higher scores on the artistic value scale. Composition, degree of
expression and aesthetic value are the three variables that have significantly higher scores of human-created artwork regardless of the attributed identity of the artist. 
However, evaluations on another variable  -- the development of personal style -- showed significant differences between conditions with AI and human-attributed artist identity. In addition, \cite{hong2019artificial} conclude that a negative attitude toward AI-created art strongly influences the evaluation of artworks when people believe they were created by AI.

The work by \cite{ragot2020ai}, in turn, shows a negative bias in the public perception towards art made by AI. In their study, paintings presented as created by humans were evaluated significantly higher in terms of four subjective dimensions: liking, beauty, novelty, and meaning. A questionnaire with Likert scales was designed for this purpose. Regarding all four variables -- declared liking, perceived beauty, novelty and meaning -- results showed the main effect of induction (AI vs. human), the type of painting (landscape vs. portrait) and the real author (AI vs. human) with small effect sizes (measured by Cohen's \textit{d}). In an additional recognition task, paintings made by humans were correctly recognised in 66\% of cases and AI-generated artwork was correctly recognised in 56\% of cases. Moreover, the recognition rate of the authors of paintings was higher for portraits than for landscapes (69\% and 53\%, respectively).

\cite{chamberlain2018putting} investigate the public response to visual art created by humans and computers.
Furthermore, the authors tested participants' ability to discriminate between computer-generated and man-made art and a potential bias towards the former category. The authors obtain similar results to \cite{ragot2020ai} in terms of prejudice and negative bias towards AI-generated art. Analysis of results reveals it is driven mostly by the belief of machines' limited abilities regarding creativity. 
The authors provide also an interesting remark regarding Turnig-like tests of intelligence. ``An important and understudied psychological question relating to this 
phenomenon is the extent to which individuals are willing to accept computer art as having 
the same worth and aesthetic value as that of a human artist, regardless of whether it passes 
such stringent tests of human-level intelligence'' \citep[p. 178]{chamberlain2018putting}.

\cite{kobis2021artificial} assessed in a TT-like fashion whether people are able to distinguish algorithm-generated versus human-written poems, as well as preference of algorithm-generated versus human-written text. The authors employed GPT-2 for the purpose of their study -- state-of-the-art Natural Language Generation algorithm \citep{radford2019language}. A text prompt comprised several lines of human-written poems that an algorithm needed to complete. There were several iterations of poem generations and either a random example (human-out-of-the-loop) or the best one (human-in-the-loop) was chosen for the study. Participants could not reliably distinguish human-written and machine-generated poems in the latter condition. However, participants slightly preferred human poems regardless of whether they were aware of the algorithmic origin of the text or not. 

In the related area of deepfakes (hyper-realistic manipulations of audio-visual content) detection, the study by \cite{kobis2021fooled} show that human participants could not reliably differentiate deepfake from authentic videos (58\% accuracy rate on average), and at the same time they overestimate their own abilities to do so. Another study found that participants had difficulty distinguishing between real and AI-generated images of people, with only 61\% able to tell the difference, far below the 85\% the researchers expected \citep{pocol2023seeing}. When justifying their classifications, 38\% of respondents cited anomalies in anatomical features (e.g., asymmetric eyes, unrealistic teeth), while 22\% referenced unnatural lighting or textures. 
This aligns with findings from \cite{} which reports a 62\% accuracy rate for AI-generated image identification among 120 participants. 
Furthermore, \citet{gu2022ai} show examples of how AI can be used to create convincing fake scientific images, that are difficult to detect even by state-of-the-art classification methods. 

The current study follows previous works in the field, presented above. Our study investigates (differences in) human judgement of aesthetic value regarding images generated by AI and humans, similarly to \citep{hong2019artificial, ragot2020ai}. However, it goes beyond these works as they focus on the investigation of fine-grained subjective dimensions of aesthetic judgements and/or influence of a reported identity of an artist on those judgements (studying prejudice/bias towards AI). On the other hand, the present study examines humans' capacity to distinguish man-made and AI-generated images with a twofold methodology. First, it follows the protocol of a parallel-paired TT (modified into a Lovelace-like TT). Second, it implements a viva voce version of TT with a single subject being examined (interrogated) at a time.
Thus, our study also compares `levels of difficulty' of different versions of TT.

\section{Methodology}
Existing studies poses at least three research gaps we aim to fulfill in the current work: (1) so far, the ability to recognised AI-generated vs. human-created art content has been examined among lay people or artists - we focus on participants with a domain knowledge, that is an educational background in cognitive and computer science; (2) studies employed either parallel-paired or viva voce version of the TT, thus the scope of evaluation and potentially introducing bias, which we show in this paper - we examine two versions of the TT, which offer distinct experimental conditions and insights regarding human cognition; (3) studies utilise prompting methods to achieve images with the desired content included, however, the quality of images generated by systems such as DALL-E can vary depending on the prompt provided often requiring multiple iterations to achieve the desired result - to this end, we select a set of existing images generated by DALL-E 2 to ensure high quality of AI-generated images, free of errors and without the confounding effects of prompt quality.
Regarding statistical analyses, we employ two-sample and one-sample t-tests to assess differences in judgements of the aesthetic value between two categories of images and differences in the recognition rate from the threshold value, respectively.

\subsection{Participants}
Data was collected from 46 Polish individuals with an educational background in cognitive science or computer science (master's students). Participants were informed about the aims of the study and the task. Informed consent was obtained from all participants prior to their participation in the study in accordance with ethical guidelines and regulations. 
Response to each item of the survey was mandatory.

\subsection{Material}
Material for the study comprises man-made paintings available on artists' public profiles (personal websites, Instagram and Pinterest accounts) and AI-generated images posted by the owner of the DALL-E 2 system -- OpenAI on platforms such as Instagram and Twitter. DALL-E 2 is used as the only source of AI-generated images.  There are, however, numerous human artists whose paintings are utilised in the study. It is currently a state-of-the-art system for image generation, similar to \cite{radford2019language}, who utilised only one language model -- GPT-2 and poems from various human writers. Therefore, an image generated by DALL-E 2 was chosen first, and then, a comparable painting in terms of style and content was selected from a human artist. This methodology allows us to regard DALL-E 2 as a machine that participates in TT.

\subsection{Survey Design}
A modified Turing Test was used to investigate the human capacity to distinguish man-made and AI-generated artworks. In addition, the study examined human judgements of the aesthetic value of artworks (images) generated by a human artist or a machine. The questionnaire comprises 12 questions in total. These questionnaire items could be grouped into three different categories of questions or three different tasks.

First, participants are asked to judge the aesthetic value of four paintings (Figure 1 in Appendix) -- two generated by AI and two created by human artists. The following four images are utilised in this part of the study: A1 - ‘Théâtre D’opéra Spatial’ AI-generated painting by Gigapixel AI; A2 - ‘Interior of the Salon of the Archduchess Isabella of Austria’ by Willem van Haecht; A3 - ‘Dino party’ generated by the OpenAI system DALL-E 2; and A4 - ‘Surrounded by color’ by Mike Winkelmann. 
The first two images were collected to match each other in terms of complexity, colour palette and tone. 
The five-point Likert scale is used for evaluating the aesthetic value of painting with option `1' meaning `I do not like this image at all' and option `5' meaning `I like this image very much'\footnote{The original question in Polish is formulated as follows: ``Na skali od 1 (nie podoba mi się) do 5 (bardzo mi się podoba), na ile podoba Ci się ten obraz?''}. 

The second type of question is a modified parallel-paired Turing Test. The Lovelace-like test of intelligence replaces the original form of the question-response pair in TT, where study participants play the role of judges (interrogators). They are presented with a pair of images -- one generated by a machine and the other created by a human. There are five pairs of images, and the participant's task is to decide which image from the pair was generated by AI\footnote{The original question in Polish is formulated as follows:``Jeden z obrazów został stworzony przez sztuczną inteligencję (ang. \textit{artificial intelligence}, AI), a drugi przez człowieka. Który obraz Twoim zdaniem został stworzony przez sztuczną inteligencję (AI)?''} (Figure 2 in Appendix). In addition, participants are asked to justify their choice in an open-question form shortly\footnote{Due to different copyright licenses, not all images are included in the manuscript file; all materials are available upon email request to the corresponding author.}.  

The following 10 images are utilised in this task: B1.a - ‘Portrait of a shiba inu astronaut, oil painting, 16th century’ AI-generated image by the OpenAI system DALL-E 2. B1.b - ‘Cat’ by Léa Roche. B2.a - ‘A dog taking a bath with bubbles in an old bathtub and smoking a cuban cigar’ generated by the OpenAI system DALL-E 2. B2.b - ‘Labradoodle’ by Lee Ann Shepard. B3.a - ‘The Kiss’ by Gustav Klimt. B3.b - Transformation of Gustav Klimt’s ‘The Kiss’ generated by the OpenAI system DALL-E 2. B4.a - ‘Donald Trump in Drag Dragrace’ by Alfredo (Argo) Rodriguez. B4.b - ‘Donald Trump’ generated by the OpenAI system DALL-E 2. B5.a - ‘A photo of an astronaut riding a horse’ generated by the OpenAI system DALL-E 2. B5.b - ‘Is-Anyone-Out-There’ by Alan Bean.

\begin{figure}[h!]
    \centering
    \includegraphics[width=0.95\linewidth]{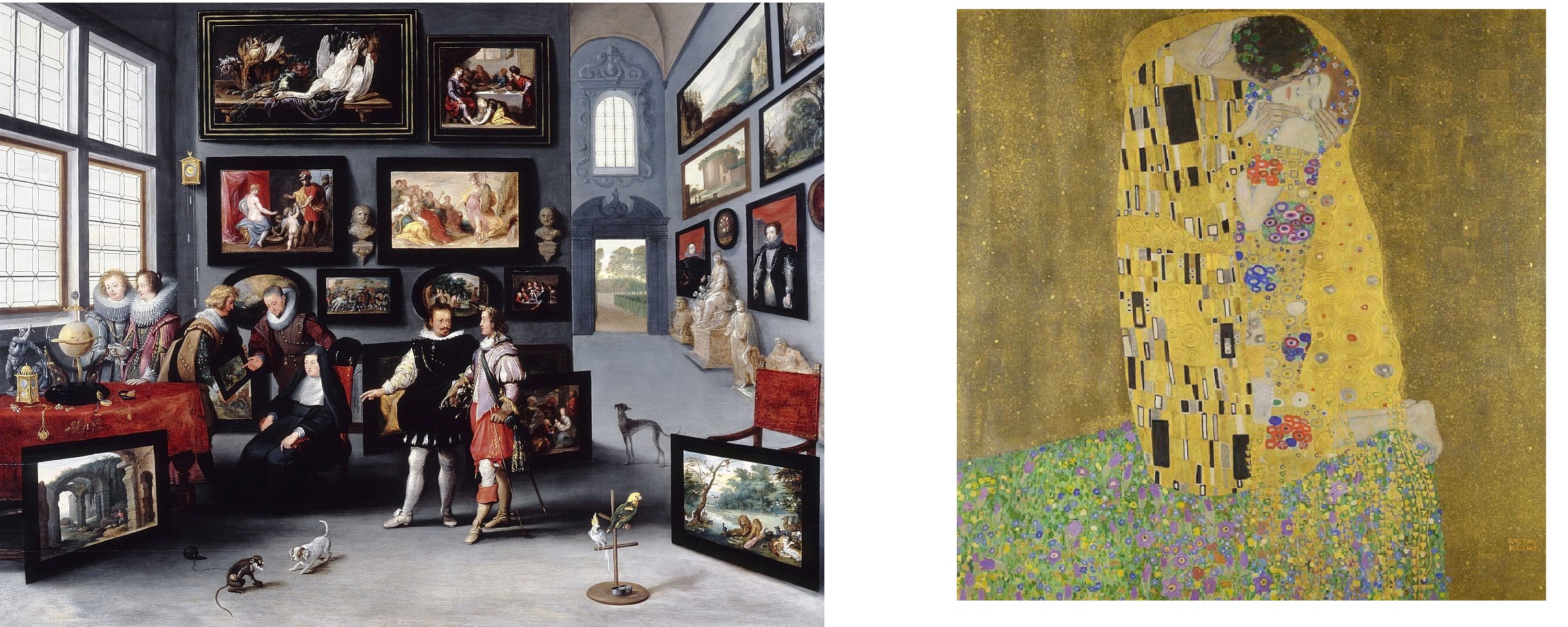}
    \caption{In the parallel-paired Turing Test, participants are presented with a pair of paintings and given the following instructions: One of the images was created by an AI program, and the other by a human artist. Which image was created by AI? Van Haecht, W. (1621). \textit{Interior of the Salon of the Archduchess Isabella of Austria} [Painting]. Mauritshuis, The Hague, Netherlands. and Klimt, G. (1908). \textit{The kiss} [Painting]. Österreichische Galerie Belvedere, Vienna, Austria.} \label{fig:pp}
\end{figure}

The third category of questions is designed in a viva voce TT fashion. Similarly to the previous type of questions, human participants play the role of judges (interrogators) in the Lovelace-like TT. However, this time, each participant is presented with a single image generated either by a human or by AI and asked whether, in their opinion, this image was generated by AI\footnote{The original question in Polish is formulated as follows: ``Czy Twoim zdaniem ten obraz został stworzony przez sztuczną inteligencję (AI)?''}. Three images are evaluated in the viva voce TT (Figure 3 in Appendix). Images chosen for the viva voce TT task include C1 - ‘Slow and steady’ by Mike Winkelmann. C2 - ‘Teddy bears mixing sparkling chemicals as mad scientists’ by the OpenAI system DALL-E 2. C3 - ‘Golden retriever puppy sitting at a diner drinking a cup of coffee, looking melancholy, edward hopper’ generated by the OpenAI system DALL-E 2.

\begin{figure}[h!]
    \centering
    \includegraphics[width=0.4\linewidth]{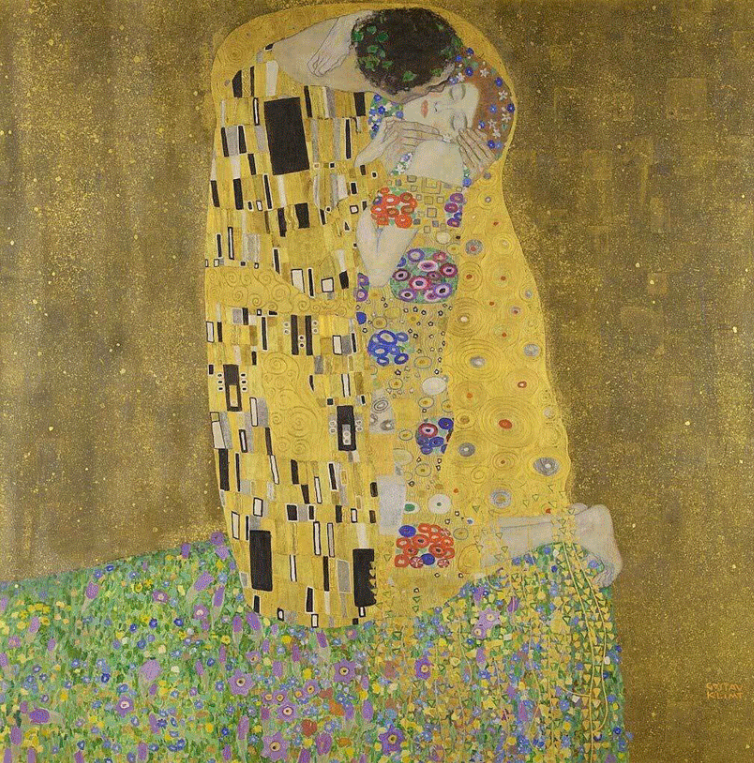}
    \caption{In the viva voce Turing Test, participants are presented with one image at a time and asked the following question: In your opinion, is this image AI-generated? Klimt, G. (1908). \textit{The kiss} [Painting]. Österreichische Galerie Belvedere, Vienna, Austria.} \label{fig:vv}
\end{figure}

\section{Results}

\subsection{Judgement of Aesthetic Value}

The first type of evaluation regards judgements of the aesthetic value of paintings generated either by AI (items A1 and A3) or a human artist (items A2 and A4). 
Table I presents a summary of the aesthetic ratings for individual images (items) and aggregated categories (AI-generated vs. human-created). The ratings were collected on a 5-point Likert scale. 
While the average ratings for AI-generated and human-created images are similar and not statistically different ($t=0.64, p=0.53$), notable differences are observed in the individual item ratings, image A1 (AI-generated) received the highest average rating among all items (3.9). In the first pair of images (A1 vs. A2), the AI-generated image is rated significantly higher ($t=3.52, p=0.001$), in the second pair (A3 vs. A4), the human-created painting is judged with a significantly higher aesthetic value ($t=-2.19, p=0.03$). 

\begin{table}[h]
    \centering
     \begin{tabular}{cccccc}
      \hline
      
      Item & M item & Mdn item &  Category & M category & Mdn category\\
      
      \hline
      
      A1 & 3.9 $\pm1.2$ & 4 &\multirow{2}{*}{AI} & \multirow{2}{*}{3.3 $\pm1.0$}  &  \multirow{2}{*}{3.5}\\ 
      
      A3 & 3.0 $\pm1.3$ & 2 &  &   \\ \hline
      
      A2 & 2.7 $\pm1.4$ & 3 & \multirow{2}{*}{Human} & \multirow{2}{*}{3.2 $\pm1.0$} & \multirow{2}{*}{3.3}\\ 
      
      A4 & 3.4 $\pm1.3$ & 4 &  & &  \\
            
      \hline
      \end{tabular}
     \caption{Summary of responses of aesthetic value of images. The average rating and the standard deviation are calculated for individual images (items) and categories (AI-generated vs. human-created).}      \label{Tab:results1}
\end{table}

\subsection{Parallel-Paired TT}
Results from this part of the survey are displayed in Table \ref{Tab:results2}. 
One-sample t-test yielded a statistically significant recognition rate above the chance level of 50\% for four pairs of images: B2, B3, B4 and B5. 
The recognition rate for pair B1 equals 50\%, that is, a random guessing of which image from the pair was created by a human and which by a machine. The overall recognition rate of 75.2\% turns out to be statistically significant above the chance level (t=8.839, p$<$0.001). 
However, given Turing's threshold of 70\% recognition rate as a measure of success in the test, only two pairs of images score significantly above this value (B3 and B4); pairs B2 and B5 are recognised at the level of 70\%, thus the machine can deceive at least 30\% of humans judges. Images in pair B1 are recognised significantly below Turing's threshold. 
Therefore, given a baseline in the form of human-created artwork, people can generally recognise AI-generated images significantly above the chance level of 50\%. On the other hand, Turing was right in his prediction that a machine will be able to deceive humans in at least 30\% of cases.

\begin{table}[h!]
    \centering
     \begin{tabular}{llll}
      \hline
      Item & Accuracy & \textit{t}-chance & \textit{t}-threshold \\
      \hline
      B1 & 50.0 & 0.0 & -2.68*\\ 
      B2 & 80.4 &  5.15***  & 1.77\\
      B3 & 93.5 &  11.81*** & 6.39***\\ 
      B4 & 87.0 &  7.36*** & 3.38**\\
      B5 & 65.2 &  2.14* & -0.68 \\ \hline
      Total & 75.2 &  8.84*** & 1.82 \\
                     
      \hline
      \end{tabular}
     \caption{Recognition rate of AI-generated images in the parallel-paired version of a modified TT. *$p<0.05$, **$p<0.01$, ***$p<0.001$.}
      \label{Tab:results2}
\end{table}

\begin{figure}[h]
    \centering
    \includegraphics[width=0.64\linewidth]{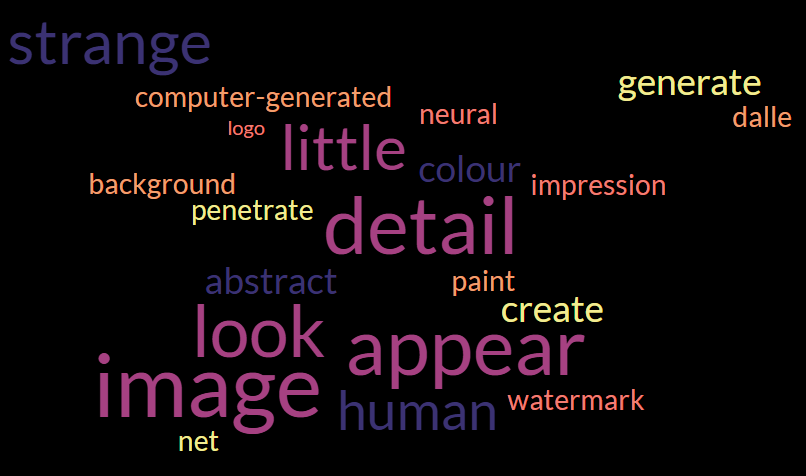}
    \caption{Qualitative analysis of open-form responses provided by participants in the parallel-paired TT.}
    \label{resqual}
\end{figure}

\paragraph{Qualitative analysis}
A qualitative (linguistic) analysis was conducted based on open-ended responses provided by the participants. In the open-ended form, participants were asked to justify their choice of an image that, in their opinion, was generated by AI. 
Textual content from these responses was normalised. That is, text was converted to lowercase; then, stop-words (function words) and punctuation marks were removed; finally, text was lemmatised, i.e. words were converted to their dictionary forms. SpaCy library, available for Python programming language, was used for this purpose. The Polish language was translated into English by the authors. 
The result of lexical analysis is depicted in Figure \ref{resqual} as maps of words (so-called word clouds).

Regarding the first pair of images (B1), participants focused on the shapes of animals painted in the artworks and the texture of the images' backgrounds. Participants who chose the correct answer (the image with a dog) referred to ``too ideal'' shapes of the astronaut dog. However, participants who chose the other image from the pair mentioned a bit of a muddled display of the image. The strange combination of a dog and the idea of an astronaut is described as both ``chaotic'' (and thus ascribed to AI) and ``creative'' (and therefore claimed to be human-generated). 
Regarding B2, participants again pointed to ``too ideal'' shapes of elements as well as the odd combination of multiple topics in the AI-generated image. 
Two groups of participants are distinguished in evaluations of the B4 pair -- the first claims image 1 (human-generated) could be created only by a human because semantic knowledge about politics and Donald Trump is required to create it; then the other group states that it is an ill-suited combination of several topics and therefore could be generated by an AI system given adequate keywords in the prompt.

\subsection{Viva Voce TT}
Results from the viva voce test are summarised in Table \ref{Tab:results3}. 
We find that participants perform no better than random guessing ($t=-0.85, p=0.40$) and significantly below the Turning's threshold ($t=-5.6, p<0.001$) in identifying AI-generated paintings in the viva voce TT. The highest recognition rate is achieved for image C2, which is significantly above the chance level of 50\% ($t=2.49, p=0.02$); however, it is not statistically different from Turing's 70\% threshold ($t=-0.37, p=0.71$). 
Image C3, in turn, was misclassified by 78\% of participants, yielding a recognition rate significantly below the chance level  ($t=-4.59, p<0.001$) and Turing's threshold ($t=-7.9, p<0.001$). 
The recognition rate for the first image C1 equals random guessing, with 50\% of participants misclassifying the image C1 as AI-generated, which is the result significantly below Turing's threshold ($t=-2.68, p=0.01$).

\begin{table}[h]
    \centering
     \begin{tabular}{llllc}
      \hline      
      Item & Accuracy & \textit{t}-chance & \textit{t}-threshold & Category \\      
      \hline      
      C1 & 50.0 & 0.0 & -2.68* & Human \\ 
      
      C2 & 67.4 & 2.50* &  -0.37 & AI   \\
      
      C3 & 21.7 & -4.60*** & -7.85*** & AI  \\ \hline
     Total & 46.4 & -0.85 & -5.55***&  - \\
              
      \hline
      \end{tabular}
     \caption{Recognition rate of AI-generated images in the viva voce version of a modified TT. *$p<0.05$, **$p<0.01$, ***$p<0.001$.}
      \label{Tab:results3}
\end{table}

\section{Discussion}
According to Turing, a machine is said to succeed in the Turing Test if a judge wrongly decides that a machine is a man and a man is a machine at least 30\% of the time \citep{turing1950computing,shah2010deception}.
Although the objective of the current study was not to test whether AI systems could pass TT explicitly, it followed the idea of TT by evaluating a machine's ability to produce creative content - the modification of the original TT suggested by Lady Lovelace. The outcomes of an AI system were assessed against those created by humans (parallel-paired TT) or the AI system per se (viva voce TT), as in Turing's original idea. Images were selected so as to best parallel each other in terms of presented topics, content and tone. Assuming that a machine is less skilled in creating art than humans, participants decided which (in B1-B5 questionnaire items) or whether (C1-C3 items) an AI-generated image falls short of a human equivalent or standard. Framing the task in this manner corresponds closely to the \cite{riedl2014lovelace}'s idea of the updated Lovelace Test. 

Regarding the viva voce version of the employed Lovelace-like Turing Test, participants did no better than random guessing in determining whether a piece of artwork (a painting) was generated by a machine or a human (46\% recognition rate). 
On the other hand, the judges could distinguish between AI-generated and man-made images in the employed parallel-paired TT significantly above the chance level (75\% recognition rate). 
Specifically, four out of five images were recognised accurately above the chance level in the parallel-paired version of the Turing Test. 
On the other hand, in two out of three cases in the viva voce version of the Turing Test, recognition accuracy does not differ from a chance level, and in one case (image C3), the accuracy of recognition falls significantly below the chance level (to 22\%). 
As a result, H1 is confirmed in the viva voce TT and rejected in the parallel-paired TT. 
In both versions of the test, however, the machine was able to deceive human judges at least 30\% of the time, fulfilling Turing's prophecy from 1950.

One may ask what are the reasons behind the discrepancy in results of these two versions of TT. It might be the case that when a judge has two items to evaluate and decide which one is generated by AI, she looks for the ``less ideal'' one. In this manner, the parallel-pair TT is easier because of this possibility of comparison between the two images (which were on purpose selected so as to present the same topic, e.g. a dog in the case of B2 and Donald Trump in the case of B4). Even participants in the open-ended form answered that they chose this specific image as AI-generated because it is ``less ... than the other'' or ``too ... than the other''. There are no specific features about these AI-generated images that make them immediately stand up as created by a machine. There are rather minor details between the two images. What is more, although participants mentioned several times that the images in pairs B2 and B5 generated by DALL-E 2 present a ``weird combination'' of ideas (e.g., an astronaut riding a horse in the cosmos), most of them still answered incorrectly that the image C3 was not generated by AI. 
Participants rated AI-generated images as 3.4 and human-generated paintings as 3.3, on average (on a 1 to 5 Likert-like scale).
As a result, H2 is confirmed -- there are no statistically significant differences in evaluations of the aesthetic value of AI-generated and man-made images. Nonetheless, results indicate differences in those judgements between individual images. 

Two main conclusions emerge from the current study that deserve further consideration. 
First, current state-of-the-art AI systems for image generation (DALL-E) can create a piece of artwork judged as aesthetically pleasing as human-created paintings. Thus, AI systems are able to achieve human-level quality in the area of image generation. 
Second, our study indicates that the result of the Turing Test -- whether or not a machine could succeed -- depends on the setup of the test (viva voce vs. parallel-paired). 
Future studies shall investigate this discrepancy with the use of different types of stimulus (for example, pictorial and linguistic samples). 
The current study could be extended further by following the \cite{riedl2014lovelace}'s idea of the updated Lovelace Test. In this setup, the DALL-E system could be provided with a proper prompt containing all the constraints specified by a judge. Then, a human (an artist or a layperson) would be provided with the same task of creating a painting, and then the judge would compare the two and tell which one satisfies the requirements. Again, a few rounds of testing would be conducted with not only different topics of images to be generated but also different judges. 
The results of this study have profound societal and ethical implications. Creativity has been considered a uniquely human trait tied to emotion, lived experiences, and cultural context. Our findings challenge this assumption by demonstrating that AI-generated art can be perceived as equally valuable as human-created works. 
As AI-generated art becomes more prevalent, ethical guidelines and policy frameworks will be necessary to address these societal challenges. Key considerations include the development of transparent authorship standards, the creation of legal protections for human artists, and the promotion of ethical AI training practices. 
First, art institutions and digital platforms could implement disclosure requirements for AI-generated content to maintain artistic integrity. Second, policymakers must establish clear copyright laws to regulate AI-generated content and prevent unfair exploitation of artists’ work.
Third, AI developers should prioritize fair data usage and consent-based training datasets to avoid plagiarism and cultural appropriation.

\section{Conclusions}

The study contributes to understanding machine intelligence through the lens of artistic creativity, leveraging human evaluation of that intelligence. By applying the Turing Test, modified to incorporate elements of the Lovelace Test, we examined whether human judges could distinguish between AI-generated and human-created artworks in different contexts: those involving three participants (parallel-paired TT) and a one-on-one interrogation (viva voce TT). 
The results showed that participants, even those with educational backgrounds in cognitive science and computer science, performed no better than chance in identifying AI-generated images under certain conditions, supporting our first hypothesis. 
Furthermore, AI-generated artworks received aesthetic evaluations comparable to those created by humans, supporting our second hypothesis that machines can produce art that resonates with human sensibilities and challenging the assumption that human creation inherently possesses qualities that machine generation cannot replicate. 
Because the survey sample consists of participants with cognitive and computer science backgrounds, results may not reflect general public perceptions, thus, further studies are needed with the extended participant pool to include laypeople and professional artists to examine whether domain expertise influences AI-art perception.

These findings challenge traditional notions of creativity as an exclusively human domain and provide empirical evidence that contemporary AI systems can generate outputs that satisfy Boden's criteria for passing a creative Turing Test — both being indistinguishable from human-created art and achieving a comparable aesthetic value \citep{boden2010turing}. 
The inability of participants to reliably distinguish between AI and human-created artworks suggests that the perceptual and cognitive mechanisms humans employ when evaluating art may rely more on surface-level features or perceived patterns rather than on deeper cognitive evaluations of origin or intention. This aligns with findings from deepfake detection studies, which similarly demonstrate limitations in human discriminatory capabilities for algorithmically generated content \citep{kobis2021fooled}. 

The fact that even participants with domain knowledge in cognitive and computer science could not reliably identify AI-generated images suggests that current systems have already achieved a remarkable level of creative sophistication. Future research shall expand on these findings by examining other creative domains beyond visual art, potentially including music and literature, to determine whether the patterns observed in this study extend across different forms of creative expression. Longitudinal studies, in turn, could track how human perception of AI-generated content evolves as these technologies become more integrated into everyday creative practices.

Finally, the study shed light on the potential for collaboration between human artists and AI systems. We show that AI can serve as a tool to augment human creativity, offering new possibilities and expanding artistic horizons. At the same time, as AI's role in society grows, it's important to consider the ethical and societal implications of its use. This includes issues of copyright, ownership, and the potential displacement of human artists. Studying these aspects ensures the responsible development and use of AI in the arts and a broader public sphere. 
Furthermore, as AI becomes more pervasive, understanding its capabilities and limitations is crucial to promote technological literacy.

\bibliography{bibT}

\end{document}